# Optimization paper production through digitalization by developing an assistance system for machine operators including quality forecast: a concept

*PREPRINT*

Moritz Schroth[1], Felix Hake[2], Konstantin Merker [3], Alexander Becher[4], Tilman Klaeger [5], Robin Huesmann[6], Detlef Eichhorn[7] and Lukas Oehm [8]

**Abstract:** *Nowadays cross-industry ranging challenges include the reduction of greenhouse gas emission and enabling a circular economy. However, the production of paper from waste paper is still a highly resource intensive task, especially in terms of energy consumption. While paper machines produce a lot of data, we have identified a lack of utilization of it and implement a concept using an operator assistance system and state-of-the-art machine learning techniques, e.g., classification, forecasting and alarm flood handling algorithms, to support daily operator tasks. Our main objective is to provide situation-specific knowledge to machine operators utilizing available data. We expect this will result in better adjusted parameters and therefore a lower footprint of the paper machines.*

## 1 Introduction

Germany's paper industry, as an elementary part of the process industry, has achieved great success in sustainability in recent years with the continuous increase in the use of waste paper. This is a big step for achieving circular economy. Paper production only using waste paper is a complex process with several production steps using a variety of additives and chemicals as well as the monitoring and adjusting machine parameters due to fluctuating waste paper qualities. The current shortage of recovered paper exacerbates these difficulties.

The process of paper production therefore requires a lot of expert knowledge. Currently operators need a lot of personal experience to produce paper in an adequate quality. But fluctuations in waste paper quality bring major uncertainties to the process stability and also results in fluctuating quality of produced paper. Using process data has great potential for harmonizing paper quality. The operator's communication about daily troubleshooting and the acquired knowledge is currently not systemized. There is a catalogue of actions, but the knowledge is stored analogue and a quick response to machine faults and downtimes is thus not possible. This problem is exacerbated by many relevant parameters to be monitored and changing input material qualities leading to a flood of alarms in the process control system (PCS) to be prioritised by operator. This can lead to many situations, especially when alarms accumulate at peak times, where no sustainable solutions to the problem is established.

[1] Fraunhofer Institute for Process Engineering and Packaging, Heidelberger Str. 20, 01159 Dresden, Germany, moritz.schroth@ivv-dd.fraunhofer.de
[2] Consulting Talents AG, Altrottstrasse 31, D-69190 Walldorf, felix.hake@consultingtalents.com
[3] Fraunhofer Institute for Process Engineering and Packaging, Heidelberger Str. 20, 01159 Dresden, Germany, konstantin.merker@ivv-dd.fraunhofer.de, 0000-0002-3395-6266
[4] University Siegen, Chair of International Production Engineering and Management, Siegener Str. 152, 57223 Kreuztal, alexander.becher@uni-siegen.de
[5] Fraunhofer Institute for Process Engineering and Packaging, Heidelberger Str. 20, 01159 Dresden, Germany, tilman.klaeger@ivv-dd.fraunhofer.de, 0000-0002-6062-7353
[6] LEIPA Group GmbH, Torgauer Str. 12-15, 10829 Berlin, robin.huesmann@leipa.com
[7] Consulting Talents AG, Altrottstrasse 31, D-69190 Walldorf, detlef.eichhorn@consultingtalents.com
[8] Fraunhofer Institute for Process Engineering and Packaging, Heidelberger Str. 20, 01159 Dresden, Germany, lukas.oehm@ivv-dd.fraunhofer.de, 0000-0002-4102-8697

Paper production is a high resource and energy consuming process. Recovered paper is the most important raw material for the paper industry. Producing one ton of paper additionally requires 17 kg process chemicals, 150 kg other additives, 9.000 l water and 2644 kWh energy. The specific $CO_2$ emission per ton of paper is 610 kg $CO_2$ on average [Ve19]. Using LEIPA as an example, process experts estimate that the use of chemicals could be reduced by 3 % and energy requirements by 5 % using intelligent data utilization. This corresponds to annual savings of approx. 750 metric tons of process chemicals, 4,800 metric tons of other additives, and 108,000 MWh of energy.

The project aims to link the data from different companies and processes within the paper circular industry and to optimize the recovery paper cycle. Therefore, a prototype AI application integrated into an operator assistance system is developed and deployed using data and implicit process expert knowledge. The envisaged solution contributes directly to the United Nations' Sustainable Development Goals (SDG), especially to SDG 9, 11, 12 and 13.

## 2 State of the art

### 2.1 Data transfer in circular economy of paper

Concepts and examples for data transfer within the paper industry value chain exist, yet they are not broadly applied. Specific data of samples taken from the paper mother reel is, in some cases, transferred analogously to packaging producers. Process optimization using process-data from paper mills is not applied.

Sorting plants aim to sort waste paper within specification of the EN643 (definitions of paper classes). Data of sorting plants is exclusively used to optimize the sorting process. Sorting process data includes Near-infrared spectroscopy (NIRS), visual data (VIS) and sometimes object recognition data. Paper mills buy paper according to the EN643, which only vaguely defines included paper fractions. The paper production process has three main leverage options to impact product quality. First is the quality of sorted paper. There are developments to support the manual inspection with NIRS and VIS. Second is the stock preparation with recovered paper being provided and pre-processed. Third option is the paper production machine. Sensor Data is collected in stock preparation and in the paper machine.

Opposing business models hinder data transfer between sorting plants and paper mills. Currently there is no benefit in selling sorted paper above specification. Mutual benefits in data transparency need to be examined. We aim to exchange data between companies and use information from sorting plants to optimize the value chain of paper.

## 2.2 Quality forecasting systems in paper production

Processes to secure product specifications are often based on laboratory measurements after production of a mother reel. Currently numerous parameters are controlled, and actions are taken when thresholds are exceeded. Selection of supervised parameters and according actions are based on knowledge of process experts. Concepts to predict machine events as i.e., web breaks are offered by companies [Ke21]. How well these systems work cannot be estimated. Prediction of stock preparation parameters using multiple linear regression models based on process and pulp data yielded promising results [Ek20]. Also, the prediction of the Canadian Standard freeness has been studied using convolutional neuronal networks [Ka21]. Neuronal networks have shown a prediction Mean Squared Error of ±8 % when used to predict laboratory quality parameters of produced paper as e.g., Scott bond and tensile strength [Ek20]. Classical mathematical models have been developed to predict relevant quality parameters of recovered paper if the composition of recycled paper is accurately described [Ka21].

Described approaches focus on the papermill as an enclosed dynamic and complex system. Yet the underlying recovered paper quality has an impact on all steps of production in a paper mill and the quality of produced paper, as it is a main resource used in the production process.

## 2.3 State estimation in machines for error detection and user assistance systems

Looking at a typical production estimating the current (error) state is one of the key factors for providing decision support. Based on the data available various options are possible. Looking at state models and comparing the outputs of the model to the actual process has been used for a long time. Here especially methods like Hidden Markov Models or timed automata have gained interest in detecting malfunctions [BN17, Vo13, Wi15, Yi00]. A more black box style approach, or phenomenological approach, as opposed to the former model-based ones can be used by applying process data directly to some kind of machine learning model with little to heavy pre-processing [Kl19, Le16]. A mixed approach querying ontologies for the process in combination with black box models is proposed in [Ra18].

Looking at data preparation of feature extraction using the process data as is (looking at a time spot) is possible, others tend to build a digital twin of the product while being build [Kl19, Sc20]. Especially for single, repeating processes analysing the data in other domains like a frequency domain is a common way of data preparation to speedup model training [Ca15, Ka19].

## 2.4 Usage of assistance systems in paper production

Beside many other industries there are many approaches in paper production, too, to support machine and process operator in their daily tasks [Fu19, Kl17 Pl17, Ra18, Sc18]. There are task differences in open systems (packaging industry) and closed systems (process industry) [MO19]. The operator tasks in paper production are in many aspects in-between. While the tasks in open systems are characterized by facing a lot of machine stops, in process industry the parameter handling is more important. Assistance systems in paper industry has to support both kinds of operator tasks. Current solutions help with different approaches for analysing and controlling the processes, store knowledge and helps to create reports. Examples are Valmet DNA Paper Reel Quality Monitoring and ABB Industrial[IT] [La08].

### 2.5 Alarm flood

In paper production a lot of alarms and warnings arrive in a control room next to the paper machine. Some of these alarms are very important for the operator and require taking action, others are simple information. Sometimes 50 % of the alarms are meaningless for the operator [Mu00]. Alarm floods can result in stress or inattention. There are different approaches to narrow down the number of incoming alarms. One approach to reduce alarm chattering is to group alarms. Another way is to find the root-cause of the alarm flood. It is also possible to find alarm sequences in historical alarm data [Ro16]. Last here mentioned approach for alarm flood filtering in process industry is Signed direct graphs (SDG) [YX12]. Most alarm flood filters are developed for chemical process industry, not for alarm flood handling in paper production. Explorative data analysis on industrial alarm and warning data combined with visualization has current and futural potential [Be19].

## 3 Concept of the machine operator assistance system

The mentioned challenges in operator tasks and the resource consuming process require a solid solution. We propose an operator assistance system (AS), that helps handling the necessary tasks, even if the employee is not a trained expert. Process knowledge is explicitly stored in the AS. Proper machine handling results in a better product quality, a reduction of web breaks, less reject and thus the required resources. Moisture reduction of 1 % in the predryer section is approximated to save 2 to 4 % of the total energy. We focus on paper production in paper circle economy as this step is the most resource consuming.

### 3.1 Concept overview

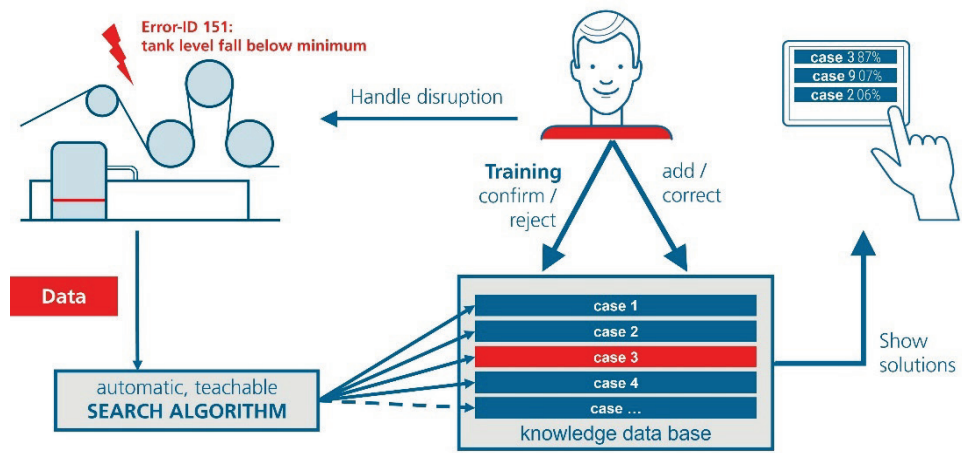

**Figure 1:** concept of the assistance system

The assistance system provides information (instructions and procedures, tools required, constraints to be observed, checklists etc.) on known events.

To save time-consuming manual searches, available process data is analysed by an intelligent search algorithm using machine learning. The trained model uses alarms and warnings from the process control system (PCS), quality forecast, and sensor data. When an event, relevant for manual examination, is detected, the assistance system suggests appropriate actions. With user feedback by means of confirmation, rejection, correction and / or supplementation, the assistance system learns with every event. The action is performed by operators by changing settings or working at the machinery. The concept of the AS is shown in figure 1. The implementation will be conducted using a microservice architecture as proposed in [Me22].

### 3.2 Knowledge base

The implicit process knowledge can be structured into components for different tasks. These components are called "knowledge cards" and are structured into three separate sections: malfunction, solution, and comment section. The malfunction section contains information on the machine error and what caused the situation. The solution section proposes different instructions how to solve the problem. The comment section can be used for further discussion or to propose changes. Because knowledge is linked, the knowledge cards can have causal relationships within the system. Therefore, if a knowledge card 'A' is the causation of knowledge card 'B' this information can be used for navigation.

To meet high quality assurance aspects, each change on a knowledge card must be approved by a card editor. Machine operators can only propose changes or discuss them in the comment section. This is critical for the correct operation of the machine as unproper actions based on wrong knowledge can cause significant costs.

### 3.3 Data sources

Many control systems are generally implemented at paper mills. Distinct control systems handle data differently and it can be a challenge to get access to these systems. In the present scenario MOPS is used as a manufacturing information warehouse system. Data is collected from many sources i.e., from the sorting process, the production process, the quality control system and laboratory data [Ek20].

The **sorting data** of the machines for the sorting process is provided via a cloud-based solution TOMRA Insight[9]. A challenge concerning the sorting data is to match it with the data of paper deliveries as there are practical limitations in tracking deliveries causing significant process overhead when implemented. So, machine learning models need to work with a non-deterministic delay from paper delivery to the sorting process.

The **process data** of the production process consists mostly of sensor data. Due to technical aspects, the data sample rate can vary and approximately every 5 to 15 seconds a new sensor value is collected for each sensor.

After a paper reel is produced, samples are taken from the production process and sent to the laboratory. Depending on the priority of a quality parameter the samples will be taken of every mother reel, every n$^{th}$ reel, or irregularly on demand. The **laboratory data** corresponding to one quality parameter has approximately up to 50 values per day.

---

[9] https://content.tomra.com/insight

The PCS raises warnings, alarms, and errors when sensor values breach operator defined thresholds. These **machine events** are used as trigger events to start querying the machine learning models.

### 3.4 Classification and forecasting algorithms

Most quality measurements of produced paper are manually taken in a laboratory after production of a mother reel. The machine operator therefore encounters a significant time delay when taking corrective actions to ensure product quality. Thus, forecasting of current paper quality is applied and shows first promising results. The goal of predicting tensile strength parameters with an accuracy of ±10 % has been reached using Random Forest Models, CatBoost and ExtraTrees Regressor. As expected, the model provides better performance when the production is running within the specifications as there are many example values to train the model. The performance decreases when predicting rare values, as for example an especially low or high value of a target variable. To achieve better results with extreme values we aim to translated to problem to a classification problem, with labels as "low", "in-specification" and "high". Using methods for anomaly detection, as the dataset for very low and high values are rare in comparison to the whole dataset, seems a promising option.

Within the project good results for the prediction of laboratory measured quality parameters as e.g., bursting strength and Scott bond have been achieved using Random Forest models. The goal is also to supervise selected parameters and to detect and inform the machine operator when system changes occur.

Presenting the best hint to the user relies on the correct identification of the current situation. Looking in the future (forecasting) the situation based in historic data can be an option. Looking at the current state is another option. As a (mostly discrete) state model is unlikely to work for a continuous process a black box model will be built using algorithms like Random Forest, proven to work in such settings [Kl19]. For the first attempt we propose to analyse a fixed state / the current process image at the very specific moment a user or an alarm triggers the assistance systems using a location identifier to provide knowledge cards. A possible extension is to have the data virtually flow and grow with the paper. In this case faults induced at a certain spot of the paper web causing faults later in the process can internally be tracked. With knowledge of the current production speed, it is therefore possible to build a digital twin of discrete parts of the web. If first approach is precise enough must be proven and heavily depends on the view of the process as a discrete or a continuous process and the time response of errors in the overall system.

### 3.5 Alarm flood filter

As the PCS of a paper machine generates many alarms, e.g., on simple excess of a single threshold, different methods for alarm flood filter will be compared and brought into action. Alarm notifications can also be reduced by only passing ones with knowledge available in the knowledge base. Here linking via error codes is applied. Another approach is to filter by smart repetition. Important notification should be repeated, but normal or not relevant notifications should not be repeated. Furthermore, patterns in alarm sequences are analysed and grouped, so that concatenated alarms are summarized in only one event.

### 3.6 Human machine interfaces

A big issue is the interaction of the AS with the operator. The operator must get all information at the correct moment. Therefore, the development of the graphical user interface (GUI) is implemented according to the Ecological Interface Design [VR92]. For visual simplifying also images and videos are used.

Knowledge to the operator is provided in one of the following situations: a web break, alarms and warnings from the PCS, recognized machine situations based on sensor and process data and predicted quality deviations. Because situation recognition and interpreting are ML-based and therefore probabilistic, the AS can only make recommendations for operators' actions. This is intended, because the AS should neither substitute operator nor inducing the operator to not thinking his- or herself. So, the operator has to make the (final) decision what to do in a certain situation.

## 4  Summary and further work

In this work we have shown the potential savings of energy, water and other resources and how we want to capitalize on them using modern techniques of the field of machine learning. We detailed how we plan to use forecasting techniques, alarm flood filtering algorithms besides other machine learning models and combine them in an assistance system for machine operators to provide situational support.

This should enable machine operators to apply fixes faster, gain a better understanding of the machine, have a structured knowledge transfer and therefore quicker learning curves. This results in higher machine uptime and a lower resources footprint, so that both environmental and economic aspects are improved.

We see the potential of further work for more creative using of data and quicker implementing of assistance systems in paper production.

**Acknowledgment:** The underlying project of this contribution was sponsored by the German Federal Ministry of Education and Research (Bundesministerium für Bildung und Forschung, BMBF), under the sponsorship reference number 02WDG019D. The authors are responsible for the contents of this contribution.